\pdfoutput=1

\documentclass[sigconf]{acmart}

\renewcommand\footnotetextcopyrightpermission[1]{} 
\usepackage[T1]{fontenc}
\usepackage[utf8]{inputenc}
\usepackage{booktabs} 
\usepackage{caption}
\usepackage{multirow}
\usepackage{graphicx}

\setcopyright{rightsretained}

\acmConference[Computation+Journalism]{Computation+Journalism Symposium}{February 2019}{Miami, USA} 
\acmYear{2019}
\copyrightyear{2019}

\hypersetup{draft}

\begin{document}
\title{Fact-checking Initiatives in Bangladesh, India, and Nepal: A Study of User Engagement and Challenges}

\author{Md Mahfuzul Haque$^1$, Mohammad Yousuf$^3$, Zahedur Arman$^4$, Md Main Uddin Rony$^2$, Ahmed Shatil Alam$^1$, Kazi Mehedi Hasan$^1$, Md Khadimul Islam$^5$, Naeemul Hassan$^2$}

\affiliation{School of Journalism and New Media, The University of Mississippi$^1$, Gaylord College of Journalism and Mass Communication, The University of Oklahoma$^3$, College of Mass Communication and Media Arts, Southern Illinois University$^4$, Department of Computer and Information Science, The University of Mississippi$^2$, Department of Political Science, The University of Mississippi$^5$}

\begin{abstract}
Fake news and misinformation spread in developing countries as fast as they do in developed countries with increasing penetration of the internet and social media. However, fighting misinformation is more difficult in developing countries where resources and necessary technologies are scarce. This study provides an understanding of the challenges various fact-checking initiatives face in three South Asian countries--Bangladesh, India, and Nepal. In-depth interviews were conducted with senior editors of six fact-checking initiatives. Challenges identified include lack of resources, technologies, and political pressure. An analysis of Facebook pages of these initiatives shows increasing user engagement with their posts.
\end{abstract}

\maketitle
\section{Introduction}
People in the digitized age constantly come across various claims often with doctored evidence ~\cite{Developmentandcooperation}. This spate of disinformation has made the fact-checking more urgent than ever. The process of fact-checking is mostly manual labor-intensive; however, we have observed in recent years that the fact-checking organizations in some developed nations have been employing automation in their verification process ~\cite{poynter}. However, fact-checkers in the developing nations do not enjoy the same facilities due to various constraints.  

The types of challenges fact-checking organizations have been facing are related to the political scene in the countries they are operating ~\cite{poynter2}, to the financial resource they have, and to the technological innovations available to them ~\cite{poynter3}. So, it can be assumed that the challenges fact-checking organizations in the developed nations have been facing are not likely to be the same in the context of developing nations, although the importance of fact-checking is no less important in these countries.

In this paper, we study five fact-checking organizations in three countries of South Asia – Bangladesh, India, and Nepal – to understand the types of challenges that are hindering their growth. Through in-depth interviews, we identified various constraints such as political pressure, lack of resource and access to technology, and lack of a sustainable business model these organizations have been facing. We also propose technological solutions these organizations can employ to make their verification process handier and more rigorous. This paper is a part of an ongoing project that focuses on the drawbacks of sustainable fact-checking organizations in the Global South.

We conducted in-depth interviews with seven fact-checkers from five organizations in three countries. In Bangladesh, three fact-checking organizations have been operating. India has six organizations, while Nepal has one. We covered all fact-checking organizations in Bangladesh and Nepal, representing 100 percent from both countries, and one from India, representing around 17 percent.

The reason for choosing these three countries in South Asia is that India is the largest democracy in the world, where spreading of disinformation has been menacing social and cultural harmony. For example, between May and June in 2018 more than a dozen people have been killed across India in violence fueled primarily by fake social media messages ~\cite{Thewashingtonpost}. In Bangladesh, disinformation on social media, especially Facebook, has been fueling political and religious instability. Crowds of angry Muslims vandalized and torched Buddhist shrines and homes in Bangladesh to protest after a fake photo of a partially burned Quran was posted on Facebook ~\cite{CNN}. The situation in Nepal is also no different. To combat the flow of disinformation, fact-checking organizations can play a pivotal role in informing people about authentic information, aiding to strengthen democratization in a society.

Studying the fact-checking organizations in these three South Asian countries helps get an in-depth knowledge of the drawbacks of the fact-checkers have been facing in doing their duties. The contribution of this exploratory study is that its findings are likely to bridge the knowledge gap among fact-checkers in South Asia and across the globe. 

\textbf{Research Questions}: Our exploration was driven by the following research questions--\textbf{RQ1:} To what extent do Facebook users engage with the fact-checking organizations of Bangladesh, India, and Nepal? and \textbf{RQ2:} What challenges do fact-checking organizations of Bangladesh, India, and Nepal face?
\vspace{-3mm}
\section{Literature Review}
With the plethora of disinformation spreading every day, there has been a rapid growth of fact-checking organizations around the world ~\cite{poynter4}. Duke Reporters’ Lab enlisted 149 fact-checking projects in 53 countries in 2018, while the number was 114 in 2017. Among these projects, around 87\% (41 of 47) fact-checkers in the United States are directly affiliated with newspapers, television networks, and other established news outlets. However, the case is different outside the U.S., where 53\% organizations (54 of 102) are directly affiliated ~\cite{poynter5}. With the growth of organizations, fact-checkers have also been facing various challenges ~\cite{poynter4}. These constraints can be categorized in the following themes: (i) political pressure, (ii) access to technology, and (iii) resource constraint.

\textbf{Fact-checking Initiatives in Bangladesh, India, and Nepal}: The history of fact-checking organizations in Bangladesh, India, and Nepal is relatively new. The organizations in Bangladesh began operating in 2017, while in India the trend is a bit older with the first known organization launched in 2014. The lone organization in Nepal started its operation in 2015. 

Bangladesh has three organizations -- BD Fact Check, Jacchai, and Fact Watch. The Fact Watch, launched in 2017, is a project run by a local university called the University of Liberal Arts. The two others have been operating independently. Jaachai and Fact Watch have their own websites, while BD Fact Check reaches audiences through its Facebook page. 

Bangladesh does not have a legal infrastructure to offer registration to the independent fact-checking organizations. As a result, these organizations have been categorized as online media outlets. These organizations mostly verify claims posted on social media relating to political, health and financial issues.

India has six organizations -- Fact Checker, Boom, Factly, Alt News, FactCrescendo, and NewsMobile. While the first four organizations have been included as “active” in Duke Reporter’s Lab, the remaining two are the members of the IFCN (International Fact-Checking Network). The organization, Fact Checker, studied in this paper is a sister concern of India Spend, a data journalism organization in India. Fact Checker mostly verifies data-driven claims. It reaches the audience through its website and social media pages. 

Nepal’s lone fact-checking organization named South Asia Check started its operation in 2015 as a non-profit venture with the finance of Panos South Asia. From its outset, the organization has been verifying claims made by politicians, ministers, bureaucrats and diplomats.

\textbf{Challenges}: Political pressure has been impeding the growth of fact-checking in the countries where press freedom is curtailed. For example, Fact-Nameh, a platform in Iran has been operated from Canada, because Iranian government did not allow the platform to work inside the country, constantly blocking its content. Operating from outside the country makes many readers doubt its credibility ~\cite{poynter2}. Fact-checkers in other countries such as China, Turkey, Zimbabwe have constantly been facing various political pressure. Fact check, a platform in China, usually verify health misinformation and cannot cover political issues because censorship is the norm in the country ~\cite{poynter2}.

Fact-checkers have been facing challenges in reaching partisan and disinterested audiences. Partisans are disinterested to accept negative conclusions about the politicians they like, and passive audiences are not interested to read something about the politicians they have never encountered ~\cite{Greenblatt2017}.

Fact-checking political issues in the developing nations has been critical because fact-checkers often come under attack on social media by supporters and opponents of the governments, leading many fact-checkers not to cover such issues. Similarly, fact-checkers in the United States have increasingly come under attack, facing accusations of “selection bias”, a term referred to the selection of topics fact-checkers choose to verify ~\cite{Kessler2018}.  

Fact-checking organizations in some countries began introducing automation in the process of verification to make the process handier. Full Fact in the United Kingdom created a tool that monitors transcripts from the BBC and debates in Parliament and identifies checkable claims; it also matches claims from an existing database of fact checks. The fact-checkers at Chequeado in Argentina created a tool named Chequeabot, which automatically scans text from 30 media outlets around the country and identifies claims from politicians; it also matches possible claims from the group’s existing database of more than 1,000 fact-checks to allow for quick reposting or tweeting of the previous fact check ~\cite{Kessler2018}.

Duke Reporters’ Lab has also developed FactStream, an app that automatically pulls fact checks from the three main U.S. fact-checkers -- FactCheck.org, PolitiFact and The Washington Post ~\cite{Adair2018}.
Brazilian fact-checking platform Aos Fatos has developed a Facebook messenger bot named “Fatima”, which automatically answers reader questions about rumors and claims ~\cite{Kessler2018}.
However, in fact-checkers in other countries such as India, Philippines, and Indonesia have been struggling to deter the spread of fake news because they do not have technological solutions to detect the spread of hoaxes on WhatsApp ~\cite{Suzuki2018}. Since it is a peer-to-peer app, fact-checker can’t get access to the messages on it and thus cannot let the people know that the messages sent to them were fake ~\cite{Faure2018}. Fact-checkers in India face another major challenge in reaching the audience due to the language barrier -- India has 23 official languages, and many more dialects ~\cite{Faure2018}.
Fact-checkers in non-English speaking countries face constraints of accessing to natural language processing tools due to its unavailability in local languages ~\cite{poynter}.

Fact-checking organizations in developing countries mostly run by the funding from the donor agencies. Due to financial constraint, organizations often cannot recruit adequate employees. For example, Bosnia-Herzegovina, a fact-checking organization in Bosnia, face constraints in verifying claims with its small-staffs ~\cite{Ricchiardi2018}. Fact-checkers often do not get access to reliable official datasets. Forty-eight fact-checkers from 4 countries gathered at the Duke University in 2018 to identify challenges in finding verifiable claims and identifying speakers from government documents. Getting funding from donors has also been challenging in some regions ~\cite{poynter}.



\section{Methods}
This study employed two methods to answer two research questions. First, the authors used a program written in Python to scrape user engagement data (e.g., number of shares, comments, reactions per post) from official Facebook pages of the fact-checking initiatives. A descriptive analysis was conducted on those data. Second, in-depth qualitative interviews with senior editors were conducted to understand challenges these initiatives have been facing.
\subsection{Facebook Data}
Using Facebook Graph API ~\footnote{https://developers.facebook.com/docs/graph-api/}, we accumulated all the Facebook posts that were created by 6 fact-checking organizations. These posts were created within May of 2014 and June of 2018. A total of 4,039 posts were collected.
A Facebook post can be a status or contain photo/video or link to an external site. We considered all type of posts for this study. For each post, we collected the status message, comments and sub-comments, and all the reactions (Like, Love, Haha, Wow, Sad, Angry) associated with a post. We used this data to examine users' engagement with the fact-checking posts. 
\subsection{In-depth Interviews}
The objective of this study is to identify the constraints that are hindering the growth of fact-checking initiatives in South Asia.
To identify the fact-checking organizations in Bangladesh, we applied snowballing sampling method to get a list of active fact-checking organizations in Bangladesh, which resulted in a total of three organizations – BD Fact Check, Jaachai, and Fact Watch. None of these organizations are enlisted in Duke Reporter’s Lab or member of IFCN, because these organizations do not fulfill the criteria to include in these networks. 
To identify the organizations in India and Nepal, we follow the lists of Duke Reporter’s Lab and IFCN, which resulted in six organizations in India -- Fact Checker, Boom, Factly, Alt News, FactCrescendo, and NewsMobile. We find one organization in Nepal named South Asia Check.     
As we interview fact-checkers from the three countries, we took approval from the Institutional Review Board (IRB). A uniform questionnaire was developed to interview fact-checkers in three countries. Upon getting approval, we piloted the questionnaire with a fact-checker from BD Fact Check in Bangladesh. After incorporating the learning from the pilot study, we developed the final questionnaire. 
To fix the appointment for an interview with the fact-checkers we communicated through email. Once the appointments were finalized, we conducted the interviews through Skype. 
We interviewed five fact-checkers from three organizations in Bangladesh -- two fact-checkers from BD Fact Check, two from Fact Watch, and one from Jaachai. 
We got a reply from Fact Checker of India and conducted an interview with its senior policy analyst. We also interviewed the editor of South Asia Check of Nepal.

\begin{figure*}[t!]
\includegraphics[width=\textwidth]{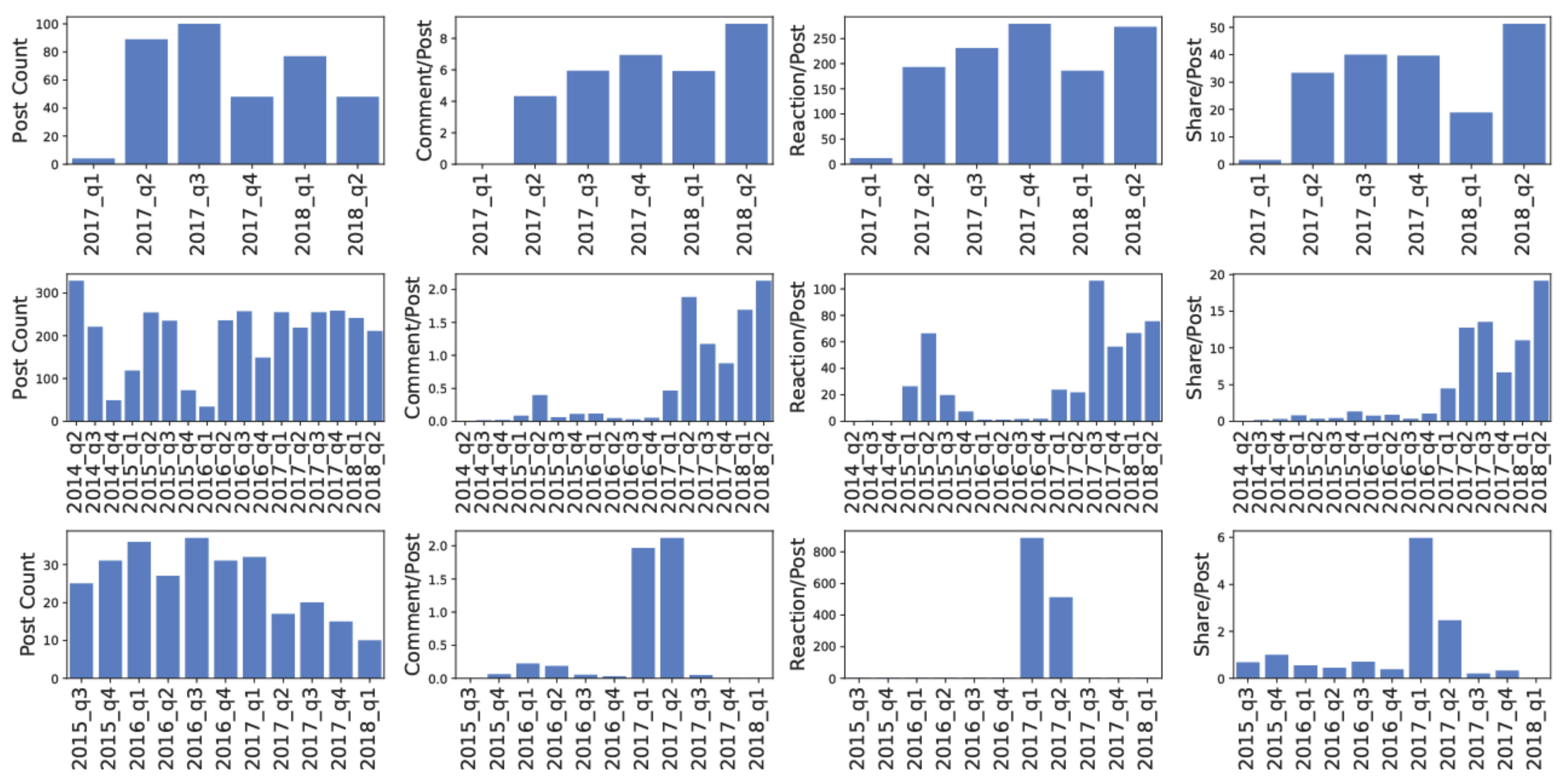}
\centering
\vspace{-5mm}
\caption{User Engagement over time for fact-checkers of three countries (Row 1 for Bangladesh, Row 2 for India, and Row 3 for Nepal)}
\label{fig:public_engagement}
\vspace{-5mm}
\end{figure*}

\section{Results}

\textbf{User Engagement}: Table ~\ref{tab:fact-check-dist-country} shows the user engagement of Facebook posts. We grouped the organizations by country. The result shows that although there was less number of posts from Bangladeshi organizations than India, they received better engagement in terms of comment, share, and reactions. Another interesting observation is that Nepali fact-checking organization had a better engagement in terms of reactions than comment and share, even greater than India. We were also interested in checking the public engagement over time. Our hypothesis is that the fact-checking organizations are getting popular day by day. Figure ~\ref{fig:public_engagement} shows the public engagement in terms of comments, shares, and reactions over time. We divided the whole time period into quarters where q1 denotes the first quarter of the year, q2 represents the second and so on. For Indian and Bangladeshi organizations, our hypothesis is mostly true (there is a decrease in the engagement at some points), but for Nepali organization, we didn't have enough evidence to prove or disapprove the hypothesis. 

\begin{table}[t!]
\centering
\resizebox{\linewidth}{!}{
\begin{tabular}{l|r|r|r|r}
\hline
\multicolumn{1}{l|}{\textbf{Country}} & \multicolumn{1}{c|}{\textbf{\#Post}} & \multicolumn{1}{c|}{\textbf{Comment Per Post}} & \multicolumn{1}{c|}{\textbf{Share per Post}} & \multicolumn{1}{c}{\textbf{Reactions Per Post}} \\ \hline
Bangladesh                            & 366                                  & 7.60                                           & 34.93                                        & 221.61                                          \\
India                                 & 3392                                 & 0.78                                           & 4.93                                         & 32.64                                           \\
Nepal                                 & 281                                  & 0.46                                           & 1.28                                         & 133.01                                          \\ \hline
\end{tabular}
}
\caption{Users' engagement with fact-checking organizations in three countries}
\vspace{-7mm}
\label{tab:fact-check-dist-country}
\end{table}

\textbf{Challenges}: Five major challenges emerged from the interviews with fact-checkers from Bangladesh, India, and Nepal. The challenges include lack of resources, lack of machine learning tools built for local languages, lack of digital archives, political pressure, and lack of a sustainable business model. Though most of these challenges persist in all three countries, political pressure remained more acute for Bangladeshi fact-checkers than others.
\subsubsection{Lack of Resources}
The fact-checking organizations under this study have been operating within serious resource constraints. All of them reported being understaffed while many can’t afford buying licensed software and, thus, rely on free tools and volunteers or part-time employees. Some organizations do not even have an office. Contributors of these organizations coordinate through the Internet. Because of lack of resources, these organizations are often forced to ignore a large number of suspicious claims.

Pointing to the need for more fact-checkers, Chaitanya Mallapur of India’s Factchecker said they “could not verify all of their queries.” Mallapur noted that some claims require sending fact-checkers to the spots where the events took place. But the lack of resources forces them to rely on government officials for information “which sometimes do not work”. Similar or worse challenges were faced by organizations in Bangladesh and Nepal. Fact-checking organizations in Bangladesh run mostly with volunteers or part-time employees. A co-founder of a Bangladeshi fact-checking organization noted, “We are a team of six members. We are all working here as volunteers”.

Naimul Karim, who works for Fact Watch in Bangladesh, said his organization did not have any funding for software and depended on free tools such as Google reverse image search and TinEye that often cannot detect fabrication if a photo is not shared publicly. Fact Watch, run by a university called ULAB and that has better funding than other organizations in Bangladesh, does not have a dedicated office. “When ULAB remains closed for any reason, we must work from home or our works stop,” Karim noted. The organization is led by three faculty members who also teach classes and have other responsibilities.

\subsubsection{Lack of Machine Learning Tools Built for Local Languages}
There are some challenges that affect all aspects of life, not only fact-checking or news organizations in particular, in the countries under study. Development of tools able to analyze texts in local languages has been extremely slow in these countries. According to Mohan Mainali, editor of the South Asia Check in Nepal, 

\begin{quote}
“If we can search for particular information from a pdf file of Nepali texts by entering keywords in electronic documents it would be easier for us to search for information we want. This facility is available in many other languages but not in the Nepali language.” 
\end{quote}

Mainali underscored the weaknesses of artificial intelligence (e.g., voice recognition, font recognition) and machine learning as they relate to developing countries. This delays the process of fact-checking. Zahed Arman from Bangladesh’s BD Fact-check echoed Mainali as he said: 

\begin{quote} 
“Western fact checking organizations are using automated fact-checking tools which is not possible in Bangladesh as there are no automated tools for Bengali language. Western fact-checking organizations are also using machine learning and artificial intelligence to identify whether a news or speech is fake or real. We don’t have such technological tools and resources to use.”
\end{quote}

Fact checkers from other organizations in Nepal and Bangladesh brought up this point while India’s Chaitanya Mallapur said his organization focuses on data, not text, and this leads to another big challenge.

\subsubsection{Lack of Digital Archives}
Fact-checkers often need to dive into archives to verify quotes or claims about past events. Having access to digital archives is a prerequisite for fast verification of these claims. But almost all of the interviewees suggested a lack of such archives in their countries. Zahed Arman noted, “There is not any tradition of storing data in the archives. We publish our report in Bengali. Thus, it’s very difficult to find out actual data in Bengali. If you find data, you cannot get access to it.”

Some government bodies in Bangladesh and India offer access to public records although they don’t have data from recent years. Mallapur from India’s Factchecker suggested differences in data collection methods by those government bodies. He clarified, “if we get data set from 2011 to 2016 then the data from every year might not be collected through the same methodological approach.” Mainali said, “Major challenge we face is lack of archive system. We have to spend a lot of time to collect information that should have been available instantly.” Though Bangladeshi fact-checkers face the same problem, some of them are optimistic. One said: 
\begin{quote}
“I think the archival system in Bengali language will improve gradually as most of the Bengali newspapers now have online archives. Most of the governmental offices have a website, and there is some information that we can use in our fact-checking purpose.”
\end{quote}
However, finding required information from these archives often remains a problem if the information is in Bengali. Several other fact-checkers from Bangladesh agreed. 

\subsubsection{Scarce Freedom of Expression}

A majority of the fact-checkers in Bangladesh suggested that they had faced pressure from public officials not to publish anything critical of the government. Some even reported receiving threats from people claiming to be senior government officials. This often leads the fact-checkers to avoid topics related to the government. Fact-checkers from India and Nepal, however, denied having such a problem. One Bangladeshi fact-checker said: 
\begin{quote}
“As a professional fact-checker, I received a direct threat from the Prime Minister Office not to publish anything that criticizes the government and its officials.”
\end{quote}
Another fact-checker mentioned pressure from both state and non-state actors: 
\begin{quote}
“When a fact check goes against a certain quarter, they become a potential threat to us in a society where democracy is too fragile, press freedom is compromised. We already have confronted more than once with state actors as BTRC have summoned us to explain our work and warned not to publish `unwanted' things. We had random phone calls from government high ups to be careful.”
\end{quote}

Some of these organizations still try to tell the truth. They often keep the identities of their contributors secret. Fact-checkers who said they did not receive such threats admitted they maintain self-censorship to make sure they don’t anger the government.

\subsubsection{Lack of A Sustainable Business Model}
On top of all the challenges described above, some of which are unique for developing countries, fact-checking organizations have one problem that exists in developed countries as well. That is the lack of a sustainable business model. Most of these organizations depend on donations and are unable to make long term plans. Mainali of Nepal said, “We are dependent mainly on donors' money which is always uncertain. We are unable to make long term plan and go ahead accordingly.” 

Fact-checkers in Bangladesh suggested that no big organization was willing to support them financially because they often challenge the narratives spread through social media by the government and the ruling party. However, donors are still the driving force of these organizations. Neither a subscription-based model nor a crowd-funded model would work in Bangladesh or Nepal. India's Factchecker, however, ``is an initiative of The Spending \& Policy Research Foundation which also runs www.indiaspend.org, India's first data journalism initiative'' (https://factchecker.in/about-us/). Mallapur of Factchecker also admitted that funding remained a major challenge for them as well. However, he believes fact-checking organizations working on public interests should remain non-profit.
Many fact-checkers in Bangladesh fear they may not be able to continue fact-checking for long because of the lack of a sustainable flow of financial resources.

One fact-checker from Bangladesh’s Jaachai noted:
\begin{quote}
“Self-funded and voluntary initiatives like ours are always at risk of getting closed down. Such initiatives should not run as a hobby or free-time job. Having a stable flow of money to keep the initiative alive and to employ necessary resources is essential.”
\end{quote}

Naimul Karim of FactWatch stressed building a model based on advertisement and public support. He also studies funding models of fact-checking organizations in the United States. Md Tajul Islam, however, seemed to disagree with his colleague as he noted that reliance on corporations may lead to biases.



\section{conclusion}
The main objective of this study is to provide a deep understanding of the challenges faced by the fact-checking initiatives in Bangladesh, India, and Nepal-three developing countries in South with a total population of over 1.5 billion. The study also looked at how users engage with these initiatives.

Independent fact-checking initiatives in these countries are part of a relatively new phenomenon that started in Western democracies in the face of a growing concern over fake news and misinformation. Most of the initiatives studied in this project lack organizational structures and operate with volunteers and part-time employees. They are different from most fact-checking organizations in Europe and the United States that are affiliated with established news outlets. Despite their limitations, these initiatives remained persistent in their pursuit of debunking false claims. Users also appear to have started to engage with these initiatives. Each initiative created a sizable community on its Facebook page. For instance, BD Fact Check, an initiative launched in Bangladesh in 2017 and run with six volunteers, has nearly 11,000 followers on its Facebook page as of October 31, 2018. The growing Facebook communities and user engagement through reaction, comment and share underscore a strong need for fact-checking organizations.

Some of the challenges identified through interviews (e.g., political pressure) are on a par with the challenges that fact-checking organizations in many other countries with similar socio-economic background have been facing ~\cite{poynter2}. Other challenges such as lack machine learning and artificial intelligence tools built for local languages and lack of searchable digital archives are specific to countries. These challenges make fact-checking more difficult in these countries than developed countries. Overall, this study stresses the need for increased attention from scholars, computer scientists, professional journalists, and investors to these countries. The challenges offer opportunities for research and development of tools to fight fake news across the world.

\bibliographystyle{ACM-Reference-Format}
\bibliography{reference} 


\begin{thebibliography}{00}


\ifx \showCODEN    \undefined \def \showCODEN     #1{\unskip}     \fi
\ifx \showDOI      \undefined \def \showDOI       #1{#1}\fi
\ifx \showISBNx    \undefined \def \showISBNx     #1{\unskip}     \fi
\ifx \showISBNxiii \undefined \def \showISBNxiii  #1{\unskip}     \fi
\ifx \showISSN     \undefined \def \showISSN      #1{\unskip}     \fi
\ifx \showLCCN     \undefined \def \showLCCN      #1{\unskip}     \fi
\ifx \shownote     \undefined \def \shownote      #1{#1}          \fi
\ifx \showarticletitle \undefined \def \showarticletitle #1{#1}   \fi
\ifx \showURL      \undefined \def \showURL       {\relax}        \fi
\providecommand\bibfield[2]{#2}
\providecommand\bibinfo[2]{#2}
\providecommand\natexlab[1]{#1}
\providecommand\showeprint[2][]{arXiv:#2}

\bibitem[\protect\citeauthoryear{Adair}{Adair}{2018}]%
        {Adair2018}
\bibfield{author}{\bibinfo{person}{Bill Adair}.} \bibinfo{year}{accessed on
  October 30, 2018}\natexlab{}.
\newblock \bibinfo{booktitle}{{\em FactStream app now shows latest fact-checks
  from Post, FactCheck.org and PolitiFact}}.
\newblock
\showURL{%
\url{https://reporterslab.org/factstream-app-now-shows-latest-fact-checks-from-post-factcheck-org-and-politifact/}}


\bibitem[\protect\citeauthoryear{Ahmed}{Ahmed}{2018}]%
        {CNN}
\bibfield{author}{\bibinfo{person}{Farid Ahmed}.} \bibinfo{year}{accessed on
  October 30, 2018}\natexlab{}.
\newblock \bibinfo{booktitle}{{\em Bangladesh Muslims torch Buddhist shrines,
  police say}}.
\newblock
\showURL{%
\url{https://www.cnn.com/2012/09/30/world/asia/bangladesh-muslim-buddhist-violence/index.html}}


\bibitem[\protect\citeauthoryear{Faure}{Faure}{2018}]%
        {Faure2018}
\bibfield{author}{\bibinfo{person}{Gaelle Faure}.} \bibinfo{year}{accessed on
  October 30, 2018}\natexlab{}.
\newblock \bibinfo{booktitle}{{\em Standing up for truth: Meet India’s 'fake
  news' fighters}}.
\newblock
\showURL{%
\url{http://observers.france24.com/en/20180413-india-fake-news-fact-checkers}}


\bibitem[\protect\citeauthoryear{Funke}{Funke}{2018a}]%
        {poynter}
\bibfield{author}{\bibinfo{person}{Daniel Funke}.} \bibinfo{year}{accessed on
  October 30, 2018}\natexlab{a}.
\newblock \bibinfo{booktitle}{{\em Automated fact-checking has come a long way.
  But it still faces significant challenges}}.
\newblock
\showURL{%
\url{https://www.poynter.org/news/automated-fact-checking-has-come-long-way-it-still-faces-significant-challenges}}


\bibitem[\protect\citeauthoryear{Funke}{Funke}{2018b}]%
        {poynter2}
\bibfield{author}{\bibinfo{person}{Daniel Funke}.} \bibinfo{year}{accessed on
  October 30, 2018}\natexlab{b}.
\newblock \bibinfo{booktitle}{{\em How to fact-check politics in countries with
  no press freedom}}.
\newblock
\showURL{%
\url{https://www.poynter.org/news/how-fact-check-politics-countries-no-press-freedom}}


\bibitem[\protect\citeauthoryear{Funke}{Funke}{2018c}]%
        {poynter3}
\bibfield{author}{\bibinfo{person}{Daniel Funke}.} \bibinfo{year}{accessed on
  October 30, 2018}\natexlab{c}.
\newblock \bibinfo{booktitle}{{\em In a step toward automation, Full Fact has
  built a live fact-checking prototype}}.
\newblock
\showURL{%
\url{https://www.poynter.org/news/step-toward-automation-full-fact-has-built-live-fact-checking-prototype}}


\bibitem[\protect\citeauthoryear{Funke}{Funke}{2018d}]%
        {poynter5}
\bibfield{author}{\bibinfo{person}{Daniel Funke}.} \bibinfo{year}{accessed on
  October 30, 2018}\natexlab{d}.
\newblock \bibinfo{booktitle}{{\em Report: There are 149 fact-checking projects
  in 53 countries. That’s a new high.}}
\newblock
\showURL{%
\url{https://www.poynter.org/news/report-there-are-149-fact-checking-projects-53-countries-thats-new-high}}


\bibitem[\protect\citeauthoryear{Gowen}{Gowen}{2018}]%
        {Thewashingtonpost}
\bibfield{author}{\bibinfo{person}{Annie Gowen}.} \bibinfo{year}{accessed on
  October 30, 2018}\natexlab{}.
\newblock \bibinfo{booktitle}{{\em As mob lynchings fueled by WhatsApp messages
  sweep India, authorities struggle to combat fake news}}.
\newblock
\showURL{%
\url{https://www.washingtonpost.com/world/asia_pacific/as-mob-lynchings-fueled-by-whatsapp-sweep-india-authorities-struggle-to-combat-fake-news/2018/07/02/683a1578-7bba-11e8-ac4e-421ef7165923_story.html?utm_term=.6f3f2a0b7722}}


\bibitem[\protect\citeauthoryear{Greenblatt}{Greenblatt}{2018}]%
        {Greenblatt2017}
\bibfield{author}{\bibinfo{person}{Alan Greenblatt}.} \bibinfo{year}{accessed
  on October 30, 2018}\natexlab{}.
\newblock \bibinfo{booktitle}{{\em How fact-checkers will respond to new
  challenges: Some solutions}}.
\newblock
\showURL{%
\url{https://www.americanpressinstitute.org/publications/reports/white-papers/fact-checkers-solutions/}}


\bibitem[\protect\citeauthoryear{Kessler}{Kessler}{2018}]%
        {Kessler2018}
\bibfield{author}{\bibinfo{person}{Glenn Kessler}.} \bibinfo{year}{accessed on
  October 30, 2018}\natexlab{}.
\newblock \bibinfo{booktitle}{{\em Rapidly expanding fact-checking movement
  faces growing pains}}.
\newblock
\showURL{%
\url{https://www.washingtonpost.com/news/fact-checker/wp/2018/06/25/rapidly-expanding-fact-checking-movement-faces-growing-pains/?utm_term=.973d2e0d78ac}}


\bibitem[\protect\citeauthoryear{Mantzarlis}{Mantzarlis}{2018}]%
        {poynter4}
\bibfield{author}{\bibinfo{person}{Alexios Mantzarlis}.}
  \bibinfo{year}{accessed on October 30, 2018}\natexlab{}.
\newblock \bibinfo{booktitle}{{\em There's been an explosion of international
  fact-checkers, but they face big challenges}}.
\newblock
\showURL{%
\url{https://www.poynter.org/news/theres-been-explosion-international-fact-checkers-they-face-big-challenges}}


\bibitem[\protect\citeauthoryear{Ricchiardi}{Ricchiardi}{2018}]%
        {Ricchiardi2018}
\bibfield{author}{\bibinfo{person}{Sherry Ricchiardi}.} \bibinfo{year}{accessed
  on October 30, 2018}\natexlab{}.
\newblock \bibinfo{booktitle}{{\em Fact-checking around the world: Inside
  Bosnia-Herzegovina’s fact-checkers}}.
\newblock
\showURL{%
\url{https://ijnet.org/en/blog/fact-checking-around-world-inside-bosnia-herzegovina’s-fact-checkers}}


\bibitem[\protect\citeauthoryear{Shiundu}{Shiundu}{2018}]%
        {Developmentandcooperation}
\bibfield{author}{\bibinfo{person}{Alphonce Shiundu}.} \bibinfo{year}{accessed
  at October 30, 2018}\natexlab{}.
\newblock \bibinfo{booktitle}{{\em The methods of fact-checking}}.
\newblock
\showURL{%
\url{https://www.dandc.eu/en/article/basic-principles-professional-fact-checking}}


\bibitem[\protect\citeauthoryear{Suzuki}{Suzuki}{2018}]%
        {Suzuki2018}
\bibfield{author}{\bibinfo{person}{Wataru Suzuki}.} \bibinfo{year}{accessed on
  October 30, 2018}\natexlab{}.
\newblock \bibinfo{booktitle}{{\em Facebook's fact-checking in Asia faces
  challenges}}.
\newblock
\showURL{%
\url{https://asia.nikkei.com/Business/Business-Trends/Facebooks-factchecking-in-Asia-faces-challenges}}


\end{thebibliography}

\end{document}